\documentclass[twocolumn,english,american]{revtex4-2}
\usepackage[T1]{fontenc}
\usepackage[utf8]{luainputenc}
\usepackage{amsmath}
\usepackage{graphicx}
\usepackage{babel}
\begin{document}
\title{Energy Window Muffin Tin Orbitals (EWMTO) and Energy Window Linear
Muffin Tin orbitals (EWLMTO) within the Atomic Sphere Approximation
(ASA)}
\author{Garry Goldstein}
\address{garrygoldsteinwinnipeg@gmail.com}
\begin{abstract}
In this work we propose two new, closely related, efficient basis
sets for the electronic structure problem. The basis sets are based
on the Muffin Tin Orbital (MTO) idea that the eigenstates of the Khon
Sham (KS) Hamiltonian may we be expanded in terms of eigenstates of
the spherically averaged KS Hamiltonian inside the so called Muffin
Tin (MT) spheres and Bessel functions in the interstitial multiplied
by appropriate spherical Harmonics. Here we use the fact that the
solution to the finding the ground state electron density is most
often found through an iterative process: where generically on the
order of over twenty iterations are taken till the ground state electron
density and energy converges to the lowest values allowed by the correlation
and exchange functional. We use eigenstate information from the previous
iteration loop to choose the energies of the basis set elements used
to study the KS Hamiltonian. Furthermore within the Atomic Sphere
Approximation (ASA) the energies of the Bessel functions do not matter,
as they are cancelled out other then for boundary conditions, and
are chosen at zero energy. This is an efficient method aimed at studying
the electronic structure of materials with large unit cells especially
if they are of close packed form where ASA is particularly accurate. 
\end{abstract}
\maketitle

\section{Introduction}\label{sec:Introduction}

In this work we study the electronic structure problem once an accurate
correlation and exchange energy $\mathcal{E}_{XC}\left(\left\{ \rho\left(\mathbf{r}\right)\right\} \right)$
has been established (here $\rho\left(\mathbf{r}\right)$ is the density
of electrons at the point $\mathbf{r}$), be it the Local Density
Approximation (LDA), Generalized Gradient Approximation (GGA), meta-GGA
or etc \citep{Martin_2020,Marx_2009,Singh_2006}. Within the Density
Functional Theory (DFT) framework in order to ascertain the ground
state energy of a crystalline solid it is important to solve the associated
Khon Sham (KS) problem given by the KS Hamiltonian \citep{Martin_2020,Marx_2009,Singh_2006}:
\begin{align}
H_{KS} & =-\frac{\nabla^{2}}{2m}+\int d^{3}\mathbf{r}'\frac{e^{2}\rho\left(\mathbf{r}'\right)}{\left|\mathbf{r}-\mathbf{r}'\right|}-\sum_{\mathbf{R}_{t}}\frac{Z_{t}e^{2}}{\left|\mathbf{R}_{t}-\mathbf{r}\right|}\nonumber \\
 & +\frac{\delta}{\delta\rho\left(\mathbf{r}\right)}\mathcal{E}_{XC}\left(\left\{ \rho\left(\mathbf{r}\right)\right\} \right)\label{eq:H_KS}
\end{align}
as accurately as possible within the limits of as small as possible
(incomplete) basis set. Here $e$ is the electron charge, $\mathbf{R}_{t}$
is the the nuclear positions (which are assumed to be fixed), $m$
is the electron mass and $Z_{t}$ is the number of protons in the
nucleus at $\mathbf{R}_{t}$. It is well known that plane waves are
a very poor approximation to eigenstates of the KS Hamiltonian near
the nuclear co-ordinates $\mathbf{R}_{t}$ (which oscillate wildly
near the nucleus and are not smooth near $\mathbf{R}_{t}$), with
impractically large plane wave basis sets are needed \citep{Martin_2020,Marx_2009,Singh_2006}.
There have been three key methods to overcome this difficulty: pseudo-wave
functions \citep{Martin_2020,Marx_2009,Blochl_1994,Vanderbilt_1990,Singh_2006},
augmented wave functions \citep{Andersen_1975,Loucks_1967,Michalicek_2013,Michalicek_2014,Singh_1991,Singh_2006,Sjostedt_2000,Slater_1937,Soler_1989,Soler_1990},
Muffin Tin Orbital (MTO) wave functions \citep{Andersen_1975,Andersen_1984,Andersen_2003,Khon_1954,Korringa_1947,Skriver_1984,Wills_2010,Goldstein_2024}.
In the case of pseudo wave functions the all electron KS problem is
replaced by another nearly equivalent KS problem where the pseudized
KS eigen wave functions are smooth everywhere including near the nuclear
co-ordinates so that the pseudized KS problem may be well solved in
terms of a basis of plane waves. In augmented wave function methods
the plane waves are augmented by solutions of some radially averaged
KS problem near the nuclei $\mathbf{R}_{t}$ (or in the so called
Muffin Tin (MT) spheres around $\mathbf{R}_{t}$ of radius $S_{t}$).
Matching conditions (in the value of the wave function and usually
its slope at the MT radius) are then imposed for an efficient basis
for the solution of the all electron KS problem. In MTO methods the
plane wave is replaced by an appropriate Bessel function multiplied
by some spherical harmonic and then matching conditions are imposed
on the Bessel function and solutions to the radially averaged KS problem.

Of the three methods, described above, MTO wave functions have the
smallest basis set for unit cell size, however they are considered
the least accurate. They are often used for large unit cells such
as those for alloys such as hardened steel where several hundred transition
$d$-orbital metal atoms are close packed inside of an atomic unit
cell. The LMTO method within the Atomic Sphere Approximation (ASA)
then uses $s$, $p$, and $d$ orbitals for each atom (or a total
of as little as nine times the total number of atoms orbitals or basis
points), allowing for hundreds of atoms with manageable computer sizes.
The main drawback of this method is that it only predicts accurate
energies in a single band, usually chosen to be the valence band (of
course multi-window options are possible where a second or even a
third linearization energy is chosen whereby it is possible to increase
accuracy, however this greatly increases computer time). Roughly speaking
the computer time scale goes as $\sim N^{3}$ where $N$ is the number
of linearization energies chosen making multi-window options often
impractical. Here we propose to have an effect similar to multi-window
options where there are several linearization energies (as much as
100 or more if need be) but the basis size does not increase with
the number of linearization energies or windows and still remains
the same size say e.g. nine (in the case of transition metals) per
atom per unit cell. The key idea is to use previous iteration loops
to assign linearization energies to approximate eigenstates of the
KS Hamiltonian and then use these as a basis for the system. We present
an explicit algorithm in Section \ref{sec:Simple-construction} below.
The construction and algorithm is related to previous works by the
author \citep{Goldstein_2024-1}.

\section{MTO and LMTO basis sets within the Atomic Sphere Approximation (ASA)
review}\label{sec:MTO-and-LMTO}

We now review the MTO and LMTO basis set within ASA. In both the MTO
and LMTO basis set the wave function consists of a sum of three components.
\begin{equation}
\chi_{t,\mathbf{k},l,m}^{\left(L\right)}\left(E,\mathbf{r}\right)=\chi_{t,\mathbf{k},l,m}^{\left(L\right)M}\left(E,\mathbf{r}\right)+\chi_{t,\mathbf{k},l,m}^{B}\left(\mathbf{r}\right)+\chi_{t,\mathbf{k},l,m}^{\left(L\right)T}\left(E,\mathbf{r}\right)\label{eq:Main_Bessel_tail}
\end{equation}
Here $\mathbf{k}$ is a wave vector in the first Brillouin zone, $\left(l,m\right)$
is the angular momentum channel, $\left(L\right)$ stands for $\left(L\right)$MTO
and $M,B,T$ stands for main, Bessel and tail components of the wave
function respectively. We now have that: 
\begin{widetext}
\begin{equation}
\chi_{t,\mathbf{k},l,m}^{\left(L\right)M}\left(E,\mathbf{r}\right)=\left\{ \begin{array}{cc}
\sum_{\mathbf{R}_{t}}\exp\left(i\mathbf{k}\cdot\mathbf{R}_{t}\right)\phi_{t,l}^{\left(L\right)}\left(E,\left|\mathbf{r}-\mathbf{R}_{t}\right|\right)Y_{lm}\left(\widehat{\mathbf{r}-\mathbf{R}_{t}}\right) & \left|\mathbf{r}-\mathbf{R}_{t}\right|\leq S_{t}\\
0 & \left|\mathbf{r}-\mathbf{R}_{t}\right|>S_{t}
\end{array}\right.\label{eq:Chi_Main}
\end{equation}
\end{widetext}

Now to obtain $\phi_{t,l,m}^{\left(L\right)}\left(E,\left|\mathbf{r}-\mathbf{R}_{t}\right|\right)$
we look for solutions of the spherically averaged KS equation:
\begin{equation}
\left[-\frac{d^{2}}{dr^{2}}+\frac{l\left(l+1\right)}{r^{2}}+\bar{V}_{KS}\left(r\right)\right]r\phi_{t,l}\left(E,r\right)=Er\phi_{t,l}\left(E,r\right)\label{eq:Secular}
\end{equation}
Where $r=\left|\mathbf{r}-\mathbf{R}_{t}\right|$. Where $\bar{V}_{KS}$
is the spherically averaged KS potential - $V_{KS}$ with: 
\begin{equation}
V_{KS}\left(\mathbf{r}\right)=\int d^{3}\mathbf{r}'\frac{e^{2}\rho\left(\mathbf{r}'\right)}{\left|\mathbf{r}-\mathbf{r}'\right|}-\sum_{\mathbf{R}_{t}}\frac{Z_{t}e^{2}}{\left|\mathbf{R}_{t}-\mathbf{r}\right|}+\frac{\delta}{\delta\rho\left(\mathbf{r}\right)}\mathcal{E}_{XC}\left(\left\{ \rho\left(\mathbf{r}\right)\right\} \right)\label{eq:KS_potnetial}
\end{equation}
Now we have that: 
\begin{equation}
\phi_{t,l}^{\left(L\right)}\left(E,\left|\mathbf{r}-\mathbf{R}_{t}\right|\right)=A_{t,l}^{\left(L\right)}\phi_{t,l}\left(E,r\right)+B_{t,l}^{\left(L\right)}\dot{\phi}_{t,l}\left(E,r\right)\label{eq:Definition}
\end{equation}
Where $\dot{\phi}_{t,l}\left(E,r\right)=\frac{\partial}{\partial E}\phi_{t,l}\left(E,r\right)$.
Where for MTO $B_{t,l}=0$. Now the coefficients $A_{t,l}^{\left(L\right)},B_{t,l}^{\left(L\right)}$
so that the wave function matches continuously or continuously differentiably
to the Neumann function for MTO and LMTO basis sets respectively.
Since we are working within ASA (where the entire volume is taken
up by MT spheres) we choose the Neumann function to be at zero energy
with:
\begin{widetext}
\[
\chi_{t,\mathbf{k},l,m}^{B}\left(\mathbf{r}\right)=\left\{ \begin{array}{cc}
\tilde{\chi}_{t,\mathbf{k},l,m}^{B}\left(\mathbf{r}\right)=\sum_{\mathbf{R}_{t}}\exp\left(i\mathbf{k}\cdot\mathbf{R}_{t}\right)n_{t,l}\left(\left|\mathbf{r}-\mathbf{R}_{t}\right|\right)Y_{lm}\left(\widehat{\mathbf{r}-\mathbf{R}_{t}}\right) & \left|\mathbf{r}-\mathbf{R}_{t'}\right|>S_{t'},\:\forall t'\\
0 & otherwise
\end{array}\right.
\]
\end{widetext}

Where $n_{t,l}\left(r\right)=\left(\frac{S_{t}}{r}\right)^{l+1}$
is the Neumann function. Now we demand that: 
\begin{align}
A_{t,l}^{\left(L\right)}\phi_{t,l}\left(E,S_{t}\right)+B_{t,l}^{\left(L\right)}\dot{\phi}_{t,l}\left(E,S_{t}\right) & =n_{t,l}\left(S_{t}\right)=1\nonumber \\
\frac{\partial}{\partial r}\left(A_{t,l}^{\left(L\right)}\phi_{t,l}\left(E,S_{t}\right)+B_{t,l}^{\left(L\right)}\dot{\phi}_{t,l}\left(E,S_{t}\right)\right) & =\frac{\partial}{\partial r}n_{t,l}\left(S_{t}\right)\nonumber \\
 & =-\frac{l+1}{S_{t}}\label{eq:demand}
\end{align}
The second equation is only for LMTO. Now we note that $\nabla^{2}\left[n_{t,l}\left(\left|\mathbf{r}-\mathbf{R}_{t}\right|\right)Y_{lm}\left(\widehat{\mathbf{r}-\mathbf{R}_{t}}\right)\right]=0$
so it may well be expanded in terms of regular Bessel functions (at
zero energy) in various MT spheres. That is: 
\begin{equation}
\tilde{\chi}_{t,\mathbf{k},l,m}^{B}\left(\mathbf{r}\right)=\sum_{l',m'}\left(\frac{\left|\mathbf{r}-\mathbf{R}_{t'}\right|}{S_{t'}}\right)^{l'}Y_{l'm'}\left(\widehat{\mathbf{r}-\mathbf{R}_{t'}}\right)P_{\mathbf{k}t,t',lm,l'm'}\label{eq:Structure_constants}
\end{equation}
Where we write $j_{t,l}\left(r\right)=\left(\frac{r}{S_{t}}\right)^{l+1}$
is the Bessel function. With this we may write: 
\begin{widetext}
\begin{equation}
\chi_{t,\mathbf{k},l,m}^{\left(L\right)T}\left(E,\mathbf{r}\right)=\left\{ \begin{array}{cc}
\sum_{\mathbf{R}_{t'}}\sum_{l'm'}\phi_{t,t',\mathbf{k},lm,l'm'}^{\left(L\right)}\left(E,\left|\mathbf{r}-\mathbf{R}_{t'}\right|\right)Y_{l'm'}\left(\widehat{\mathbf{r}-\mathbf{R}_{t'}}\right) & \left|\mathbf{r}-\mathbf{R}_{t'}\right|\leq S_{t'}\\
0 & \left|\mathbf{r}-\mathbf{R}_{t'}\right|>S_{t'}
\end{array}\right.\label{eq:Wavefunction}
\end{equation}
Where we insure the continuity and differentiability (for MTO and
LMTO wave functions respectively) of the total wave function through
the expansion: 
\begin{equation}
\phi_{t,t',lm,l'm'}^{\left(L\right)}\left(E,\left|\mathbf{r}-\mathbf{R}_{t'}\right|\right)=A_{t,t',\mathbf{k},lm,l'm'}^{\left(L\right)}\phi_{t',l'}\left(E,r\right)+B_{t,t',\mathbf{k},lm,l'm'}^{\left(L\right)}\dot{\phi}_{t',l'}\left(E,r\right)\label{eq:Defintion}
\end{equation}
Where: 
\begin{align}
A_{t,t',\mathbf{k},lm,l'm'}^{\left(L\right)}\phi_{t',l'}\left(E,S_{t'}\right)+B_{t,t',\mathbf{k},lm,l'm'}^{\left(L\right)}\dot{\phi}_{t,l}\left(E,S_{t'}\right) & =j_{t',l'}\left(S_{t'}\right)P_{\mathbf{k}t,t',lm,l'm'}=P_{\mathbf{k}t,t',lm,l'm'}\nonumber \\
\frac{\partial}{\partial r}\left(A_{t,t',\mathbf{k},lm,l'm'}^{\left(L\right)}\phi_{t',l'}\left(E,S_{t'}\right)+B_{t,t',\mathbf{k},lm,l'm'}^{\left(L\right)}\dot{\phi}_{t',l'}\left(E,S_{t'}\right)\right) & =\frac{\partial}{\partial r}j_{t,l}\left(S_{t}\right)P_{\mathbf{k}t,t',lm,l'm'}=\frac{l}{S_{t}}P_{\mathbf{k}t,t',lm,l'm'}\label{eq:Demand-1}
\end{align}
Where the second equation is only for LMTO.
\end{widetext}

\section{Main construction}\label{sec:Simple-construction}

The main construction is based on an iterative construction of a basis
set based on the fact that it takes multiple (on the order of twenty
or more) iterations for the KS equations to converge. We will describe
initialization propagation and looping of the basis set. 

\subsection{Initialization}\label{subsec:Intialization}

We initialize the basis set in the form of $\chi_{t,\mathbf{k},l,m}^{\left(L\right)}\left(E,\mathbf{r}\right)$
for some initial electron density $\rho\left(\mathbf{r}\right)$ we
then setup the matrices and some linearization energy $E$. We then
compute:
\begin{align}
H_{tlm}^{t'l'm'}\left(\mathbf{k}\right) & =\left\langle \chi_{t,\mathbf{k},l,m}^{\left(L\right)}\left(E\right)\right|\left[-\frac{\nabla^{2}}{2m}+V_{KS}\right]\left|\chi_{t',\mathbf{k},l',m'}^{\left(L\right)}\left(E\right)\right\rangle \nonumber \\
O_{tlm}^{t'l'm'}\left(\mathbf{k}\right) & =\left\langle \chi_{t,\mathbf{k},l,m}^{\left(L\right)}\left(E\right)\mid\chi_{t',\mathbf{k},l',m'}^{\left(L\right)}\left(E\right)\right\rangle \label{eq:Main_matricies}
\end{align}
We now solve for the eigenvalue problem: 
\begin{equation}
\sum_{t'l'm'}H_{tlm}^{t'l'm'}\left(\mathbf{k}\right)\kappa_{t'l'm'}^{\varepsilon}=\varepsilon\left(\mathbf{k}\right)\sum_{t'l'm'}O_{tlm}^{t'l'm'}\left(\mathbf{k}\right)\kappa_{t'l'm'}^{\varepsilon}\left(\mathbf{k}\right)\label{eq:Secular-1}
\end{equation}
Whereby we can obtain the electron density and as such the new KS
potential $V_{KS}$. This procedure sets up the self consistency cycle
as we have obtained a list $\left\{ \varepsilon\left(\mathbf{k}\right),\kappa_{tlm}^{\varepsilon}\left(\mathbf{k}\right)\right\} $. 

\subsection{Energy Windows and looping}\label{subsec:Energy-Windows-and}

We start with a list $\left\{ \varepsilon\left(\mathbf{k}\right),\kappa_{tlm}^{\varepsilon}\left(\mathbf{k}\right)\right\} $.
We now introduce a set of energy windows where we associate to a set
of $\left\{ E_{W}\right\} $ energy sectors between:
\begin{equation}
E_{W}^{Low}\leq\varepsilon<E_{W}^{Up}\label{eq:Window}
\end{equation}
 so that $E_{W}\left(\varepsilon\right)$ is the energy corresponding
to the energy window in Eq. (\ref{eq:Window}). We now choose the
basis: 
\begin{equation}
\chi_{\mathbf{k}}^{\left(L\right)}\left(\varepsilon\right)=\sum_{tlm}\kappa_{tlm}^{\varepsilon}\left(\mathbf{k}\right)\chi_{t,\mathbf{k},l,m}^{\left(L\right)}\left(E_{W}\left(\varepsilon\left(\mathbf{k}\right)\right)\right)\label{eq:Basis}
\end{equation}
We now compute:
\begin{widetext}
\begin{align}
H_{\varepsilon}^{\varepsilon'}\left(\mathbf{k}\right) & =\left\langle \chi_{\mathbf{k}}^{\left(L\right)}\left(\varepsilon\right)\right|\left[-\frac{\nabla^{2}}{2m}+V_{KS}\right]\left|\chi_{\mathbf{k}}^{\left(L\right)}\left(\varepsilon'\right)\right\rangle \nonumber \\
 & =\sum_{tlm}\kappa_{tlm}^{\varepsilon}\left(\mathbf{k}\right)\sum_{t'l'm'}\left\langle \chi_{t,\mathbf{k},l,m}^{\left(L\right)}\left(E_{W}\left(\varepsilon\right)\right)\right|\left[-\frac{\nabla^{2}}{2m}+V_{KS}\right]\left|\chi_{t',\mathbf{k},l',m'}^{\left(L\right)}\left(E_{W}\left(\varepsilon'\right)\right)\right\rangle \kappa_{t'l'm'}^{\varepsilon'}\left(\mathbf{k}\right)\nonumber \\
O_{\varepsilon}^{\varepsilon'}\left(\mathbf{k}\right) & =\left\langle \chi_{\mathbf{k}}^{\left(L\right)}\left(\varepsilon\right)\mid\chi_{\mathbf{k}}^{\left(L\right)}\left(\varepsilon'\right)\right\rangle \nonumber \\
 & =\sum_{tlm}\kappa_{tlm}^{\varepsilon}\left(\mathbf{k}\right)\sum_{t'l'm'}\left\langle \chi_{t,\mathbf{k},l,m}^{\left(L\right)}\left(E_{W}\left(\varepsilon\right)\right)\mid\chi_{t',\mathbf{k},l',m'}^{\left(L\right)}\left(E_{W}\left(\varepsilon'\right)\right)\right\rangle \kappa_{t'l'm'}^{\varepsilon'}\left(\mathbf{k}\right)\label{eq:Main_secular_matricies}
\end{align}
\begin{figure}
\selectlanguage{english}%
\begin{centering}
\includegraphics[width=1\columnwidth]{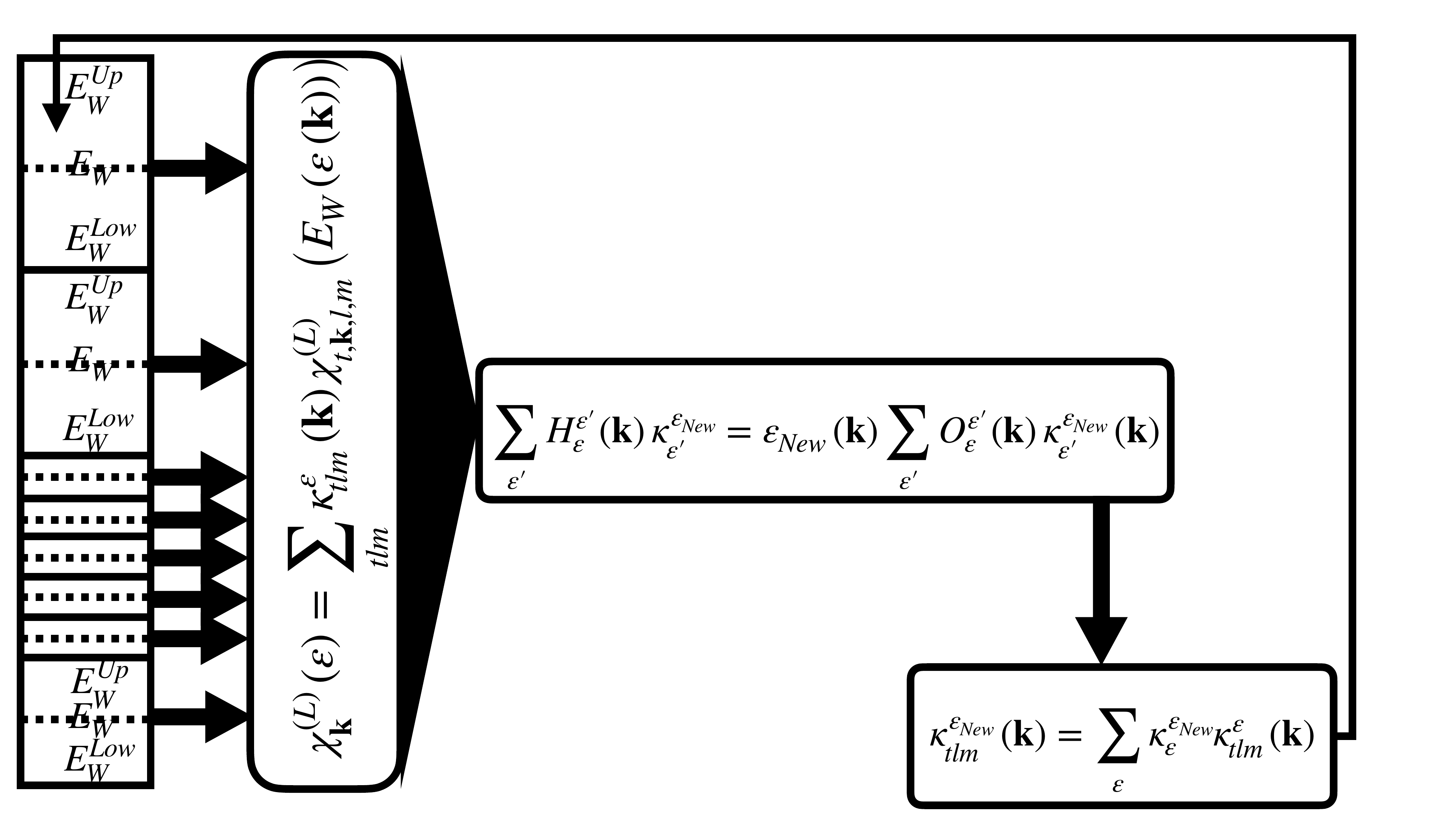}
\par\end{centering}
\caption{The coding flow consists of four main stages.}\label{fig-Coding_flow}
\selectlanguage{american}%
\end{figure}
\end{widetext}

We now form the secular equation

\begin{equation}
\sum_{\varepsilon'}H_{\varepsilon}^{\varepsilon'}\left(\mathbf{k}\right)\kappa_{\varepsilon'}^{\varepsilon_{New}}=\varepsilon_{New}\left(\mathbf{k}\right)\sum_{\varepsilon'}O_{\varepsilon}^{\varepsilon'}\left(\mathbf{k}\right)\kappa_{\varepsilon'}^{\varepsilon_{New}}\left(\mathbf{k}\right)\label{eq:Secular-2}
\end{equation}
We now form: 
\begin{equation}
\kappa_{tlm}^{\varepsilon_{New}}\left(\mathbf{k}\right)=\sum_{\varepsilon}\kappa_{\varepsilon}^{\varepsilon_{New}}\kappa_{tlm}^{\varepsilon}\left(\mathbf{k}\right)\label{eq:Coefficient_matrices}
\end{equation}
As such we have obtained the needed initial data set $\left\{ \varepsilon^{New}\left(\mathbf{k}\right),\kappa_{tlm}^{\varepsilon_{New}}\left(\mathbf{k}\right)\right\} $.
We now drop the name $New$ and start again with the same problem
till convergence. This is illustrated in Fig. (\ref{fig-Coding_flow}).
We see that the construction flows and converges to the solution of
the KS problem. 

\section{Conclusions}\label{sec:Conclusions}

In this work we have considered a new type of LMTO and MTO basis.
The idea was to expand the flexibility of the basis set to include
more energy dependence without extending the basis set, say to multiple
copies at various energies, or being dependent on a self consistent
condition like the Khon, Korringa, Rostoker (KKR) construction \citep{Khon_1954,Korringa_1947}
- where the energy is chosen self consistently say using tail cancellation.
These procedures are numerically expensive. We have done so by exploiting
the fact that solution to the self consistency cycle for the KS problem
consists of many (on the order of twenty) iterative cycles where the
secular equation for the current KS Hamiltonian is solved. Information
from the previous solution of the KS Hamiltonian in the form of eigenvalues
and eigenvectors is then used to construct the basis set for the next
iteration of the self consistency loop. This allows for a basis set
that is energy dependent without having to solve for a self consistent
energy equation. We presented two closely related basis sets of MTO
and LMTO type. Given that the basis is energy adaptive the MTO construction
is sufficient and likely better then the LMTO one. The author expects
many applications of this basis set procedure to real materials in
the near future. This work is similar in nature to the work of the
author \citep{Goldstein_2024-1} however it deals with MTO based basis
sets rather then APW ones.
\selectlanguage{english}%


\begin{thebibliography}{10 (2013)}
\bibitem[1(2006)]{Singh_2006} D. J. Singh and D. Nordstrom, \textit{Planewaves,
pseudopotentials, and the LAPW method} (Springer, New York, 2006).

\bibitem[2(2020)]{Martin_2020} R. M. Martin, \textit{Electronic Structures
Basic Theory and Practical Methods} (Cambridge University Press, Cambridge,
2020).

\bibitem[3(2009)]{Marx_2009} D. Marx and J. Hutter, \textit{Ab Initio
Molecular Dynamics Basic Theory and Advanced Methods} (Cambridge University
Press, Cambridge, 2009).

\selectlanguage{american}%
\bibitem[4(1990)]{Vanderbilt_1990}\foreignlanguage{english}{ D. Vanderbilt,
Phys. Rev. B \textbf{41}, 7892 (1990).}

\bibitem[5(1994)]{Blochl_1994}\foreignlanguage{english}{ P. E. Blochl,
Phys. Rev. B \textbf{50}, 17953 (1994).}

\selectlanguage{english}%
\bibitem[6(1975)]{Andersen_1975}O. K. Andersen, Phys. Rev. B \textbf{12},
3060 (1975).

\bibitem[7(1989)]{Soler_1989} J. M. Soler, and A. R. Williams, Phys.
Rev. B\textbf{ 40}, 1560 (1989).

\bibitem[8(1990)]{Soler_1990} J. M. Soler and A. R. Williams, Phys.
Rev. B \textbf{42}, 9728 (1990).

\bibitem[9(2014)]{Michalicek_2014}G. Michalicek, \textit{Extending
the precision and efficiency of all-electron full-potential linearized
augment plane-wave density functional theory} (Aachen University,
2014, thesis).

\bibitem[10(2013)]{Michalicek_2013} G. Michalicek, M. Betzinger,
C. Freidrich and S. Blugel, Comp. Phys. Comm. \textbf{184}, 2670 (2013).

\bibitem[11(1991)]{Singh_1991} D. J. Singh, Phys. Rev. B \textbf{43},
6388 (1991).

\bibitem[12(2000)]{Sjostedt_2000} E. Sjostedt, L. Nordstrom, and
D. J. Singh, Sol. Sta. Comm. \textbf{114}, 15 (2000).

\bibitem[13(1937)]{Slater_1937} J. C. Slater, Phys. Rev. \textbf{51},
846 (1937).

\bibitem[14(1967)]{Loucks_1967} T. L. Loucks, \textit{Augmented Plane
Wave Method} (W. A. Benjamin Inc., New York, 1967).

\bibitem[15(2010)]{Wills_2010} J. M. Wills, M. Alouani, P. Anderson,
A. Dellin, O. Eriksson, and O. Grechnyev, \textit{Full-Potential Electronic
Structure Method Energy and Force Calculations with Density Functional
Theory and Dynamical Mean Field Theory} (Springer, New York, 2010).

\bibitem[16(1984)]{Skriver_1984} H. L. Skriver, \textit{The LMTO
method Muffin-Tin Orbitals and Electronic Structure} (Springer, New
York, 1984).

\bibitem[17(1984)]{Andersen_1984} O. K. Andersen, and O. Jepsen,
Phys. Rev. Lett. \textbf{53}, 2571 (1984).

\bibitem[18(2003)]{Andersen_2003} O. K. Andersen, T. S.-Dasgupta,
and S. Ezhof, Bull. Mat. Sci. \textbf{26}, 19 (2003).

\bibitem[19(1954)]{Khon_1954} W. Kohn, N. Rostoker, Phys. Rev. \textbf{94},
1111 (1954).

\bibitem[20(1947)]{Korringa_1947}J. Korringa, Physica \textbf{13},
392 (1947).

\selectlanguage{american}%
\bibitem[21(2024)]{Goldstein_2024}\foreignlanguage{english}{ G. Goldstein,
arXiv 2403.12846.}

\bibitem[22(2024)]{Goldstein_2024-1}\foreignlanguage{english}{ G.
Goldstein, arXiv 2405.11926}

\end{thebibliography}
\end{document}